# Two-dimensional skyrmions and other solitonic structures in confinement-frustrated chiral nematics

Paul J. Ackerman,[1,2] Rahul P. Trivedi,[1,2] Bohdan Senyuk,[1] Jao van de Lagemaat,[1,3,5] and Ivan I. Smalyukh[1,2,4,5,*]

[1]*Department of Physics, University of Colorado, Boulder, Colorado 80309, USA*
[2]*Department of Electrical, Computer and Energy Engineering, University of Colorado, Boulder, Colorado 80309, USA*
[3]*National Renewable Energy Laboratory, Golden, Colorado 80401, USA*
[4]*Liquid Crystal Materials Research Center and Materials Science and Engineering Program, University of Colorado, Boulder, Colorado 80309, USA*
[5]*Renewable and Sustainable Energy Institute, National Renewable Energy Laboratory and University of Colorado, Boulder, Colorado 80309, USA*

*Email: ivan.smalyukh@colorado.edu

**Abstract**

We explore spatially localized solitonic configurations of a director field, generated using optical realignment and laser-induced heating, in frustrated chiral nematic liquid crystals confined between substrates with perpendicular surface anchoring. We demonstrate that, in addition to recently studied torons and Hopf-fibration solitonic structures (hopfions), one can generate a host of other axially symmetric stable and metastable director field configurations where local twist is matched to the surface boundary conditions through introduction of point defects and loops of singular and nonsingular disclinations. The experimentally demonstrated structures include the so-called "baby-skyrmions" in the form of double twist cylinders oriented perpendicular to the confining substrates where their double twist field configuration is matched to the perpendicular boundary conditions by loops of twist disclinations. We also generate complex textures with arbitrarily large skyrmion numbers. A simple back-of-the-envelope theoretical analysis based on free energy considerations and the nonpolar nature of chiral nematics provides insights into the long-term stability and diversity of these inter-related solitonic field configurations, including different types of torons, cholesteric-finger loops, two-dimensional skyrmions, and more complex structures comprised of torons, hopfions, and various disclination loops that are experimentally observed in a confinement-frustrated chiral nematic system.

# I. INTRODUCTION

Localized topological field configurations, such as skyrmions, baby-skyrmions (also referred to as two-dimensional skyrmions), and merons, play important roles in many branches of modern science [1–19]. They are two-dimensional (2D) and three-dimensional (3D) field configurations with nontrivial global topological structure (i.e., similar to point and line defects, these configurations cannot be deformed into a uniform field by a smooth or continuous process). Within the last two decades, baby-skyrmions have been found to play an important role not only in aspects of particle physics and astrophysics, but also in condensed matter and nonlinear physics [2–16]. Although some types of localized topologically interesting structures have been recently observed experimentally in chiral magnets and in liquid crystal (LC) systems such as cholesteric blue phases subjected to strong external fields [11–15], various other theoretically predicted types of related topologically interesting textures still remain elusive [7–9]. One of the reasons for this elusiveness is the effect of surfaces and confinement, which can perturb these textures and make them energetically unstable. Interestingly, under certain conditions, confinement may instead stabilize arrays of two-dimensional skyrmionlike structures [7]. It is therefore of great interest to explore how localized configurations, such as baby-skyrmions, terminate to match various boundary conditions at confining surfaces, which is one of the key goals of our study.

A large number of localized structures are known to occur spontaneously in confinement-frustrated chiral nematic liquid crystals when perpendicular surface boundary conditions compete with the chiral medium's tendency to twist, giving rise to various linear and axially symmetric structures that locally introduce twisted configurations [20]. Although at least four different types of localized linear structures, the so-called "cholesteric fingers," have been identified and successfully modeled numerically [20–22], the axially symmetric twist configurations remain much less understood [23–31]. Polarizing optical microscopy (POM) [23–27], historically the most common technique used to study LC textures, often yields similar textural appearance for rather different intricate LC field configurations [28] on account of its lack of 3D resolution [32,33]. Thus, past experimental studies of this system attempted to suggest a single structural model to explain these axially symmetric structures commonly referred to under a single umbrella term as "cholesteric bubbles" (also often called "spherulitic domains") [8,23–27,34–48]. More recently, different types of laser beams were used to generate localized structures with well-defined nontrivial topology, such as three different types of torons [28], cholesteric finger loops [33], and Hopf fibration structures (hopfions) [29]. The basic element of both torons and hopfions is a double twist torus (i.e., a loop of a double twist cylinder with a twist of the director from the cylinder axis in all looped cylinder's radial directions) [28,29]. The nematic ground state manifold $\mathbf{R}P^2$ is a sphere with antipodal points identified, and these two- and three-dimensional skyrmions (double twist cylinders and hopfions) are labeled by elements of the second and third homotopy groups, $\pi_2(\mathbf{R}P^2) = \mathbf{Z}$ and $\pi_3(\mathbf{R}P^2) = \mathbf{Z}$, respectively. Although the three different toron structures contain additional point defects or line defect loops, torons and hopfions are homeomorphic to each other as the additional point and disclination ring defects that occur in torons topologically self-compensate each other and can annihilate [28,29]. In principle, as well as being optically generated in confinement-frustrated chiral nematics, these localized twist structures (such as torons, finger loops, and hopfions) can potentially also occur naturally during quenching of cholesteric samples from isotropic phase to the chiral nematic mesophase, prompted by different hydrodynamic instabilities, flows, etc. Thus, these recently studied laser-induced localized configurations could be several of many subclasses of the structural family that were referred to in the past as "cholesteric bubbles" or "spherulitic domains." The present study explores field configurations of several other types of axially symmetric 2D structural solitons that can also be realized in confined cholesteric LCs and also represent some of the many subclasses of cholesteric bubbles naturally occurring or generated by various means.

In this work we use confined cholesteric LCs as a model system for understanding the interplay of topology of localized twisted structures and boundary conditions. We generate various axially symmetric localized textures by tightly focused laser beams and then study how the ensuing twisted director configurations vary across the sample thickness and terminate at confining substrates with the help of two different 3D director imaging techniques, combined with the conventional POM. This allows us to explore how the director field $\hat{n}(\mathbf{r})$ of localized structures can be matched to surface boundary conditions, the study of which can be extended to the understanding of skyrmionic configurations in other condensed matter systems. Chiral nematic LCs are ideally suited for this task of a model system because our recently developed techniques for 3D optical imaging have enabled high-resolution 3D visualization of director fields [32,33]. We demonstrate that both singular and nonsingular defect loops and point defects can arise as a result of confining locally twisted solitonic structures within cells with strong surface boundary conditions and only small perturbations to the director configuration exist in cells with weak surface anchoring. Our study also shows that naturally occurring cholesteric bubbles can correspond to a host of topologically different two-dimensional field configurations with twist occurring predominantly in radial directions (and not across the cell thickness) accompanied by both singular and nonsingular defects [49]. These field configurations are clearly distinguishable from types of torons and hopfions that we studied previously [28,29]. One new example from the family of such structures is the so-called baby-skyrmion that emerges as a double twist cylinder oriented perpendicular to cell substrates. In addition to axially symmetric structures with the lateral dimensions on the order of cholesteric pitch, we also demonstrate the feasibility of generating configurations of various nested finger loops. The midplane cross sections of these structures can be characterized by the so-called skyrmion number $N_s$, a topological characteristic commonly used to describe 2D skyrmions in other research fields [50]. This characteristic allows us to classify different observed structures and also to relate them to their known analogs in other condensed matter systems. In our case, $N_s$ describes the topology of the liquid crystal director field of the 2D localized texture when the nonpolar director tangent to the average local orientation of rodlike molecules in the cell's midplane is "vectorized," i.e., decorated by a vector field $\mathbf{n}(\mathbf{r}) = \mathbf{n}(x,y)$. This skyrmion topological number is defined by

$$N_s = \frac{1}{4\pi} \iint d^2\mathbf{r}\, \mathbf{n} \left( \frac{\partial \mathbf{n}}{\partial x} \frac{\partial \mathbf{n}}{\partial y} \right)$$

and counts how many times the vectorized $\mathbf{n}(\mathbf{r})$ wraps the unit sphere. A unique feature of the field configurations discussed in this work is that some of the localized structures display high values of Ns, which so far has not been achieved in chiral magnets or other condensed matter systems. Furthermore, the surface anchoring interactions and the LC medium's ability to host both surface and bulk half-integer disclination loops that, due to its nonpolar nature, in addition to surface and bulk point defects, highly enriched diversity of field configurations can be realized. Therefore, for a given value of $N_s$, one can typically realize several different configurations that differ from each other because they possess different singular and nonsingular defects as well as surface-induced perturbations of the bulk solitonic configurations.

These findings establish chiral nematic LCs as a platform for experimental exploration of solitonic field configurations that have universal importance, ranging from particle physics to condensed matter, nonlinear physics, and cosmology.

## II. MATERIALS AND TECHNIQUES

The studied chiral nematic LC cells were assembled using glass plates with transparent indium tin oxide (ITO) electrodes. To obtain strong perpendicular boundary conditions at the inner surfaces of the glass substrates, we spin coated them with polyimide JALS-204 (obtained from JSR, Japan), which was then cross linked by baking at about 220°C for 1 h to form thin alignment layers. In addition, perpendicular boundary conditions were also induced by molecular monolayers obtained by dip coating glass substrates in aqueous solutions of surfactants such as 1 wt.% [3-(trimethoxysilyl) propyl]octadecyl-dimethylammonium chloride (DMOAP) or cetyltrimethylammonium bromide (CTAB) for relatively weak perpendicular surface anchoring. The LC cell gap thickness was set using either glass spacers or strips of Mylar film placed along the cell edges. The cell gap thickness ($d$) was varied within 5–25 μm and measured after cell assembly with a spectrophotometer using interference or, alternatively, from confocal and nonlinear optical microscope images of the cell's vertical cross sections. LC cells with uniform thickness as well as wedge-shaped cells (the dihedral angle between the plates was kept below 3 deg, so that the local thickness gradients in our cells can be neglected) with varying thickness were filled with cholesteric materials of different pitch ($p$). The LC mixtures were infused into cells by capillary forces when heated to their isotropic phase, which allowed us to avoid the uncontrolled effects of flow on the alignment.

Chiral nematic LCs were prepared by doping nematic hosts MLC-6815, ZLI-3412, and E7 having positive and ZLI-2806 and MLC-6609 having negative dielectric anisotropy with a right-handed (CB15) or left-handed (ZLI-811) chiral additive (all materials obtained from EM Industries). Adding different concentrations of the chiral agents controlled the length of the cholesteric pitch. The observation of similar localized structures in cholesteric systems with different material parameters, such as elastic and dielectric constants, indicates that they are robust with respect to varying such parameters and are not material specific. For 3D imaging using fluorescence confocal polarizing microscopy (FCPM), the LCs were additionally doped with dye N,N′-Bis(2,5-di-tert-butylphenyl)-3,4,9,10-perylenedicarboximide (BTBP, obtained from Sigma-Aldrich). Excitation of this dye was achieved by an Ar-ion laser at 488 nm while fluorescence detection utilized interference filters to select fluorescence signals in the range 510–550 nm. Alternatively, 3D imaging by use of three-photon excitation fluorescence polarizing microscopy (3PEF-PM) utilized self-fluorescence signals from mesogens, such as pentylcyanobiphenyl, and chiral additives like CB15, which were excited by means of three-photon absorption. In this latter case, the self-fluorescence signals were detected within the spectral range of 380–450 nm.

The partially polymerizable cholesteric LC composite was prepared by first mixing 69% of nonreactive nematic mixture E7 with 30% of a diacrylate nematic (consisting of 12% of RM 82 and 18% of RM 257 in the final mixture) and 1% photoinitiator Irgacure 184 (from CIBA Specialty Chemicals), which was then followed by doping this nematic mixture with CB15 to obtain a cholesteric of pitch equal to 8 μm [30]. The nematic compounds have been obtained from Merck and EM Chemicals. The ensuing mixture was first dissolved in dichloromethane to homogenize, heated to 85°C for one day to remove the solvent through slow evaporation, and cooled down to obtain a room temperature chiral nematic mixture with the equilibrium cholesteric pitch of interest.

The integrated system comprising the capabilities of simultaneous optical generation of localized structures, POM, and 3D optical imaging is built around an inverted microscope IX 81 (Olympus). The laser manipulation part of the setup utilizes a reflective, electrically addressed, phase-only spatial light modulator (SLM) obtained from Boulder Nonlinear Systems (XY series, P512-1064). The SLM has 512×512 pixels, each 15×15 $\mu m^2$ in size. This setup employs an ytterbium-doped fiber laser (YLR-10-1064, IPG Photonics) operating at 1064 nm. The laser beam is linearly polarized with a Glan-laser polarizer and the polarization direction is adjusted with a half-wave retardation plate to optimize the phase modulation efficiency of the SLM. Before the beam is incident on the SLM, it is expanded to overfill the active area of the SLM; upon reflection from the SLM, it is resized again so as to overfill the back aperture of the objective. The SLM controls the phase of the beam on a pixel-by-pixel basis according to the computer-generated holographic patterns supplied at a refresh rate of 30 Hz for the entire pixel array, allowing one to use both regular Gaussian beams and also optical vortices (e.g., Laguerre-Gaussian beams). This spatially phase-modulated beam is imaged at the back aperture of the microscope objective, which generates the user determined 3D spatial trap pattern in the sample.

A dichroic mirror DM-IR (Chroma Technology Corp.) reflects the infrared trapping beam at 1064 nm while allowing visible light (used for POM imaging and excitation/detection of fluorescence for 3D imaging) to transmit through it to a charge-coupled device (CCD) camera or photomultiplier tube (PMT). For FCPM imaging, we use a laser-scanning fluorescence confocal unit (FV300, Olympus), which scans the excitation laser beam (Ar laser, 488 nm) with galvano mirrors. For 3PEF-PM imaging, we have employed a tunable (680–1080 nm) Ti-sapphire oscillator (Chameleon Ultra II, Coherent) emitting 140 fs pulses at the repetition rate of 80MHz, also scanned with the same galvano mirrors. The 3PEF-PM signal from the LC is collected in epi-detection mode with a photomultiplier tube (H5784-20, Hamamatsu) and a series of interference filters. We use a 100× oil-immersion objective with high numerical aperture NA = 1.4 (obtained from Olympus). The same objective is used for both the optical generation as well as imaging of the localized structures of interest. The vertical position of the focal point within the sample is adjusted using a stepper motor mount for the objective, with 10 nm precision. 3D images are reconstructed using signals originating from individual pixels by means of Fluoview computer software (Olympus). The integrated setup for holographic laser manipulation and 3D imaging is described in more detail elsewhere [33].

The equilibrium $\hat{n}(\mathbf{r})$ structures of axially symmetric finger loops without singularities were obtained by means of numerical minimization of elastic free energy using the director relaxation method, as described in detail elsewhere [28] (note that experimentally we also observe various configurations with singular disclination loops, which cannot be modeled using this method and will be explored numerically in more detail elsewhere). The computer-simulated POM textures corresponding to these $\hat{n}(\mathbf{r})$ structures were then obtained using a Jones matrix approach implemented in Mathematica (from Wolfram Research) and also using the experimental material and cell parameters [31]. To utilize the Jones matrix method, we discretize the cell into a stack of thin sublayers (parallel to the confining substrates) such that it is valid to assume that the orientation of $\hat{n}(\mathbf{r})$ is constant across the thickness of one sublayer. Each simulated POM image was then obtained as a result of successive multiplication of Jones matrices corresponding to the polarizer, a series of thin LC slabs each equivalent to a phase retardation plate, and the analyzer. Performing these calculations for each pixel and composing a two-dimensional texture with coordinate-dependent POM intensity analogous to the experimental images then yielded a computer-simulated POM image. Computer-simulated vertical 3PEF-PM and FCPM cross sections were obtained by first finding the coordinate-dependent angles $\theta(\mathbf{r})$ between $\hat{n}(\mathbf{r})$ and the linear polarization of the probing laser light and then plotting the normalized signal intensity as $I_{\text{3PEF-PM}} = \cos^6\theta$ for 3PEF-PM (unpolarized detection) and as $I_{\text{FCPM}} = \cos^4\theta$ for FCPM (polarized detection) [31–33]. The effects of finite resolution as well as defocusing and polarization changes due to the medium's birefringence were not accounted for as the artifact is mitigated by the use of LCs with low birefringence as well as the partial photopolymerization approach (allowing for an order-of magnitude reduction of the effective birefringence upon the replacement of the unpolymerized component of the system with an immersion oil [30]). In addition to being helpful from the standpoint of artifact-free imaging, this photopolymerization of cholesteric structures also increases the stability of certain types of structures with large values of skyrmion number $N_s$, as we discuss below.

## III. RESULTS

### A. Optical generation of elementary baby-skyrmionlike cholesteric structures

In addition to torons [28,30,31] and hopfions [29] that locally relieve boundary-condition-imposed twist frustration in chiral nematic LCs, we have discovered multiple other localized configurations (or solitons), which also enable a similar frustration relief via either spontaneous occurrence or when induced by a laser beam. Similar to torons and hopfions, they too are obtained in a cell with frustrated geometry, i.e., in a cell with homeotropic surface anchoring conditions on the confining substrates of a chiral nematic LC. The initial uniform director configuration contains only a vertical $z$ component (with the $z$ axis being normal to cell substrates) of the director field, i.e., $n_z = 1$, $n_x = 0$, and $n_y = 0$. To observe or generate the twisted solitonic structures, the thickness of the cell was varied roughly within $(0.5–1)p$. Despite the general similarities with conditions for generation of torons and hopfions, one parameter that (when varied) allows for achieving a large number of new localized configurations is the strength of surface anchoring on the confining substrates. Torons and hopfions generally require strong homeotropic boundary conditions on both substrates [28,29], so as to support the existence of point defects or disclination rings and/or strong elastic distortions in the LC bulk. Soft anchoring conditions promote generation of several other localized structures, some of which (although not all) cannot be realized in the case of strong perpendicular surface anchoring. These structures are imaged in 3D using FCPM and 3PEF-PM and classified based on their skyrmion number and singular defects accompanying the twist configurations. From the in-plane and vertical cross-sectional images, as well as 3Dimages processed using ParaView software (from Kitware Inc.), we deduce their internal configurations and present schematic representations of the reconstructed molecular director field $\hat{n}(\mathbf{r})$ as well as describe the topology of constituent defects.

Axially symmetric structures shown in Fig. 1 are localized excitations in the form of double twist cylinders with $\pi$ twist from their axis in all radial directions, matching their homeotropic exterior. It can be identified as a baby-skyrmion (i.e., a two-dimensional skyrmion), similar to that observed in magnetic and other condensed matter systems [2–10]. The double twist cylinder terminates on the confining surfaces and the director structure is matched with the vertical surface boundary conditions by small loops of singular twist disclinations, thus allowing the frustration-relieving structure to be compatible with the homeotropic anchoring at the substrates. The details of the director structure of the looped twist disclination close to a homeotropic surface are depicted in Fig. 2. The vertical extent of the double twist cylinder is practically the entire thickness of the cell $d$, approximately equal to $p$, while laterally it is somewhat narrower, having width only slightly larger than $0.5d$, as seen from the vertical cross-sectional images shown in Figs. 1(b)–1(d). Although most of such structures have double twist cylinder axes orthogonal to confining plates [Fig. 1(b)], about 20% of them have these axes at small oblique angles with respect to the plates [Figs. 1(c) and 1(d)]. The schematic reconstruction of $\hat{n}(\mathbf{r})$ in the vertical plane shows the location of twist disclinations (red rings and points depicting smaller defect loops). The structure can occasionally occur spontaneously after quenching the sample from an isotropic to cholesteric phase and can be reliably generated optically with the help of a focused Laguerre-Gaussian beam with a topological charge $l = \pm(0\text{-}4)$ and optical power up to 100 mW (at the sample). During generation, the beam is typically focused at (or close to) the cell substrate. Tight focusing of such a high-intensity beam may induce local heating (especially in cells containing indium tin oxide electrodes which are strongly absorbing at the wavelength of the trapping beam, 1064 nm) and partially affect the surface anchoring, making it easier for defects to be induced at the surfaces. Figure 3 shows a related structure, which also can be identified as a baby-skyrmion because it also is a double twist cylinder with a $\pi$ twist in all radial directions. The difference in this case is that, because of the relatively weak surface anchoring, the twist disclination loop is not well defined within the bulk of the LC or at the LC-substrate interfaces, as seen in the vertical cross section in Fig. 3(a) and depicted in the schematic representation in Fig. 3(b). In addition to being localized at the LC-solid substrate interfaces, these twist disclinations can be virtual, i.e., outside of the physical dimensions of the cell. A salient feature differentiating the structures shown in Figs. 1–3 from torons and Hopf fibrations is that the defects comprising them can be pinned to the surface or can be virtual, and not necessarily in the LC bulk. This requires strong deviations from the easy axis orientation at the confining substrates, and hence the structures with surface-bound and virtual defects are realizable only in the presence of relatively weak homeotropic boundary conditions. Since the director twists by $\pi$ in radial directions of structures shown in Figs. 1–3, they can be characterized by elementary skyrmion numbers $|N_s| = 1$.

### B. Individual loops of cholesteric finger with $2\pi$ twist

Figure 4 shows a different kind of optically generated axially symmetric localized structure with a $2\pi$ twist in all in-plane radial directions ($|N_s| = 2$). Its elaborate 3D configuration of $\hat{n}(\mathbf{r})$ is lacking mirror symmetry with respect to the cell midplane. The structure has a lateral expanse larger than its vertical dimension. It is generated with the help of LG beams having high $l = \pm(12$–$20)$ and optical power of about 100 mW, while using cells with indium tin oxide electrodes promoting local melting due to light absorption. From the in-plane and vertical cross sections, such as the ones shown in Figs. 4(a) and 4(b), we reconstruct the $\hat{n}(\mathbf{r})$ for this axially symmetric configuration, as depicted in Fig. 4(c). The reconstructed director field configuration contains two loops of nonsingular $\lambda$ disclinations of opposite winding numbers, situated in the LC bulk (loops of $\lambda^{+1/2}$ and $\lambda^{-1/2}$ nonsingular defect lines are marked on the schematic). The structure is anchored to the substrate opposite from the location of nonsingular defects via two loops of surface-bound singular twist disclinations, the inner one of which, in this case, has shrunk to a point defect (or a small loop that cannot be resolved due to finite optical resolution). In the lateral plane, the director turns by total of $4\pi$ in the middle plane of the cell, as can be seen from Fig. 4(c). The two nonsingular and two singular disclinations ensure conservation of the global topological charge as the twist is embedded into a cell with a uniformly aligned vertical far-field director. Two more examples of axially symmetric configurations that embed a twist of $2\pi$ from their vertical symmetry axes to the periphery in all radial directions are shown in Figs. 5 and 6 (they also can be characterized by $|N_s| = 2$). Based on 3PEF-PM images shown in Figs. 5(a)–5(c) we have reconstructed the director field configuration shown in Fig. 5(d) that also contains two half-integer $\lambda$ disclinations of opposite winding numbers and two twist disclinations, all forming circular loops. Unlike the case of twist configurations with radially outward axes of twist (the so-called helical $\chi$ axis) roughly parallel to the plane of the cell (Figs. 1–5), the axially symmetric localized excitation shown in Fig. 6 has a helical axis tilted with respect to the cell substrates. It contains nonsingular half-integer $\lambda$ disclinations in the bulk and singular line and point (or small ring-like) defects at the confining surfaces (Fig. 6).

The solitonic structures described above are inter-related and can be understood by drawing analogies with linear, translationally invariant defect structures found in similar cell geometries, viz., the so-called cholesteric fingers [20]. For example, the axially symmetric structure shown in Fig. 1 resembles its linear counterpart, the so-called cholesteric finger of the third type, CF3 [20]. The translationally invariant structure of the CF3 has a net twist of $\pi$ in its midplane and the in-plane helical axis perpendicular to the finger (and hence is rather narrow in width). This twist of CF3 is matched to the perpendicular boundary conditions with the help of twist disclinations on opposite surfaces of the cell. Thus, the axially symmetric structure shown in Fig. 1 can be thought of as an axially symmetric loop of CF3. Similarly, the structures shown in Figs. 4 and 5 can be understood as loops of the cholesteric finger of the fourth type, CF4, which has a $2\pi$ twist along a helical axis perpendicular to the finger and contains two twist disclinations at one surface and two nonsingular $\lambda$ disclinations in the bulk closer to the other confining surface. Since the CF4 finger is not self-mirror symmetric with respect to the vertical plane along its length, there are two ways of forming a loop of CF4 by choosing interior and exterior sides of the loop, which correspond to the two structures shown in Figs. 4 and 5. These structures are generated by focusing the laser beam near one substrate with an ITO electrode and causing local heating and/or perturbations to alignment on one cell substrate's surface.

The structure shown in Fig. 6 can be identified as a loop of cholesteric finger of the first kind, CF1, having the helical axis tilted with respect to the cell normal, $2\pi$ twist along this helical axis, and a quadrupole of nonsingular half-integer $\lambda$ disclinations of opposite signs forming loops and embedding this twist into a cell with vertical boundary conditions. Once a loop of such a CF1 finger is formed, relatively weak surface boundary conditions allow for some of the defect lines of the quadrupoles to transform into surface disclinations and some to collapse into a surface boojum or small subresolution disclination loops (marked by a blue filled circle in the schematic). This structure is not the only way of forming a loop of CF1. We have reported imaging and reconstruction of different mechanism of looping of CF1 finger configurations in the past in Ref. [33] and also show two more examples in Fig. 7. In Figs. 7(a) and 7(b) one can see a version of a CF1 loop in which one of the $\lambda^{-1/2}$ disclinations has shrunk into a point defect [as seen in the bottom of the vertical cross-section 3 in Fig. 7(a)]. On the contrary, all four $\lambda$ disclinations of CF1 finger loop shown in Figs. 7(c)–7(j) stay nonsingular and are less clearly defined within the twisted structure. These configurations are obtained by various forms of dragging the laser beam focused close to the center of the cell, along a circle to connect the laser-induced CF1 finger into a loop, which then shrinks to its equilibrium size upon turning off the laser light. There is no local heating involved in this type of generation of the structure depicted in Figs. 7(c)–7(j). However, the transformation of a finger loop like the one shown in Figs. 7(c)–7(j) into a structure with a point defect, as the one shown in Figs. 7(a) and 7(b), is possible through laser-induced heating of the sample with the CF1 finger loop at one of the confining substrates, whereby the nonsingular disclination loop shrinks into a point defect upon quenching the sample locally back to the mesophase.

### C. Nested axially symmetric structures of multiple torons and finger loops

Although the structures described above may potentially occur spontaneously as metastable or stable configurations under the influence of quenching from isotropic to chiral nematic phase, flows, and various fields applied using patterned electrodes, our exquisite optical control with tightly focused beams also achieves configurations that cannot be accessed otherwise. Examples of such axially symmetric structures that are highly unlikely to occur spontaneously are shown in Fig. 8. We generated these structures with laser beams of 75 mW and higher by first generating a toron, such as the ones seen in the centers of configurations shown in Figs. 8(a) and 8(d). Then we "massaged" the toron with a laser beam by steering the beam along a circular path in the cell midplane with a diameter slightly larger than the size of the toron. This laser manipulation at the periphery of the toron forces the structure to "split" into a loop of a CF3 finger having two twist disclinations close to the opposite confining substrates of the cell (note that a somewhat different type of laser "massaging" described in Ref. [29] was instead transforming torons into hopfions). The loop of CF3 can be resized by continuously adjusting the diameter of the circle along which the laser beam is being dragged. By increasing the size of the ensuing CF3 loop, we create a homeotropic region in its interior that is large enough to host additional torons or finger loops within it. This procedure was repeated to obtain concentric structures with many levels of nested finger loops and torons (Fig. 8). The structures could also be obtained in a LC system that can be partially polymerized [Figs. 8(a), 8(d), 8(f), and 8(g)], allowing for its increased stability and feasibility of imaging such structures by 3PEF-PM while avoiding some of the artifacts, as we discussed in detail earlier [30]. On the other hand, in regular nonpolymerized LCs, the use of relatively high laser powers of 80 mW and higher allows for pinning of the CF3 loop configurations at desired locations [Figs. 8(b) and 8(c)], thus increasing the number of possible field configurations that can be obtained by means of laser manipulation, including concentric axially symmetric and many other structures. For example, the pinning of the exterior loop of CF3 in Fig. 8(c) not only allows one to control its shape (e.g., making it rectangular), but also leaves ample room in the homeotropic interior within which, in principle, one could generate arbitrary configurations of loops and torons—concentric as well as eccentric, with a single or many different centers or other configurations of nested fingers and torons. Surface pinning of the finger loops or/and partial polymerization are often (although not always) essential for the stability of such overall complex topological structures. For example, when generated at relatively low laser powers of 50–70 mW (at which no surface pinning of the loops is observed), if the structure shown in Fig. 8(d) were left unpolymerized, the toron in the center would get squeezed and disappear, and the immediate outer loop would transform back into a toron and so on until there would be only one stable toron left. However, photopolymerization [30] and/or surface pinning of nested finger loops allow for the long-term stability of such complex topological configurations. 3D imaging [Figs. 8(f) and 8(g)] reveals that the interior structure of these complex configurations is a toron, which is surrounded by the loop of CF3 [Figs. 8(d) and 8(f)] or by many such CF3 loops [Figs. 8(b) and 8(c)], as discussed below.

## IV. DISCUSSION

To obtain insights into the long-term stability and diversity of the studied inter-related solitonic field configurations, one can perform a simple back-of-the-envelope theoretical analysis based on free energy considerations and the nonpolar nature of chiral nematic LCs. The Frank elastic free energy cost of different spatially localized structures occurring in chiral nematic LCs can be calculated as [49]

$$F_{\text{elastic}} = \int \left\{ \frac{K_{11}}{2}(\nabla \cdot \hat{n})^2 + \frac{K_{22}}{2}\left[\hat{n} \cdot (\nabla \times \hat{n}) + \frac{2\pi}{p}\right]^2 + \frac{K_{33}}{2}\left[\hat{n} \times (\nabla \times \hat{n})\right]^2 \right.$$
$$\left. - K_{24}\nabla \cdot \left[\hat{n}(\nabla \cdot \hat{n}) + \hat{n} \times (\nabla \times \hat{n})\right] \right\} dV, \tag{1}$$

where $K_{11}$, $K_{22}$, $K_{33}$, and $K_{24}$ are elastic constants describing splay, twist, bend, and saddle splay distortions of $\hat{n}(\mathbf{r})$, respectively. We will assume that the role of the saddle splay term can be neglected. In a uniform unwound cholesteric LC confined between glass plates with strong perpendicular boundary conditions, the elastic free energy density cost (per area $A$) of completely unwinding the helix of the cholesteric ground state is $F_{\text{elastic}}/A = 2\pi^2 K_{22}d/p^2$. For a typical axially symmetric structure with lateral dimensions comparable to $p$ and typical cell thickness $d = 10$ µm, one finds $F_{\text{elastic}} \sim 10^{-15}$ J, much larger than thermal energy ($\sim 10^5 \, k_B T$). To understand the stability and diversity of observed solitonic structures, it is instructive to compare this free energy to the energetic cost of various defect loops. Nonsingular disclinations are expected to be much less energetically costly than the singular defect lines, and we therefore first estimate the free energy of the latter to show that different defects, including singular disclination loops, can be stabilized by locally introduced twisted director configurations. Neglecting the difference in the values of Frank elastic constants, the line tension (free energy per unit length) of such a singular disclination can be roughly estimated as

$$T_{\text{defect}} = \frac{\pi}{4} K \ln \frac{L}{r_{\text{core}}} + T_{\text{core}}, \tag{2}$$

where $K$ is the average elastic constant, $L \sim d$ is the size of the system, and $r_{\text{core}}$ and $T_{\text{core}}$ are the radius and energy per unit length of the defect core. Assuming an isotropic (melted) core model [49], one can estimate the core energy as $T_{\text{core}} \approx \pi k_B \Delta T_{\text{NI}} \rho N r_c^2 / M$, where $M$ is the molecular mass, $k_B$ is the Boltzmann constant, $N$ is the Avogadro's number, $\Delta T_{\text{NI}}$ is the difference between the temperature of the transition from LC to the isotropic phase and the room temperature (at which all experiments were done), and $\rho$ is the density of the LC. By minimizing the overall elastic and core free energy, one finds the core radius $r_{\text{core}} \approx 10$ nm. Taking typical cell and material parameters of thermotropic LCs, such as $K = 10$ pN, one then finds $T_{\text{defect}} \sim 50$ pN. The relative costs of introducing loops of singular defect lines and lowering of the elastic free energy through enabling local twist are the two main competing factors that determine relative stability of our laser-generated structures, such as the ones shown in Figs. 1–5. Indeed, the free energy of a singular defect loop of lateral size equal $p$ is roughly $\pi p T_{\text{defect}} \sim 10^{-15}$ J, comparable to the energetic cost of unwinding the cholesteric LC by the perpendicular surface boundary conditions. Although highly simplified, the above estimates are consistent with our experimental observations, which show that the loops of singular and nonsingular defect lines can mediate formation of various metastable-state twist configurations to locally relieve frustration in the unwound cholesteric LC (Figs. 1–6). The structures with nonsingular defect lines alone have defect loops of negligible energetic cost because of their twist-escaped core structures, but they are accompanied with considerable bend and splay distortions of the director, which also can be of energetic cost comparable to that of singular defect loops. Indeed, the uniform homeotropic textures have no contributions to the splay and bend terms in Eq. (1) but have a large (costly) twist energy term. The laser-induced solitonic twisted structures locally minimize the twist term of $F_{\text{elastic}}$ at the expense of introducing additional splay and bend distortions and introducing point defects and defect loops with singular or nonsingular cores. For $d \approx p$, contributions of different terms in Eq. (1) are such that the solitonic structures have elastic energy comparable to that of the homeotropic unwound state, but the two states are separated by elastic energy barriers much larger than the thermal energy. These large energetic barriers are responsible for the long-term stability or metastability and for the selective generation of a host of different twisted solitonic structures. With the approach of laser generation demonstrated here, one can cross these energy barriers via optical realignment at laser powers in the range of 50–100 mW (measured at the sample plane). This is driven by minimization of the free energy term that describes coupling to the electric field of the generating Laguerre-Gaussian laser light ($\mathbf{E}_{\text{LG}}$) and $\hat{n}(\mathbf{r})$, which is proportional to $\int d^3\mathbf{r}(\mathbf{E}_{\text{LG}} \cdot \hat{n})^2$ and increases linearly with laser light intensity. In addition to the optical realignment of $\hat{n}(\mathbf{r})$, laser light focused at or close to substrates with indium tin oxide electrodes prompts heating and even a localized phase transition to the isotropic phase, allowing one to further guide the localized director field configurations, especially the ones with surface defects.

The above estimates, because of their qualitative nature, only justify the existence and explain diversity of the solitonic structures in chiral nematics. The main goal of this simple analytical exercise is to show that it is the chiral term of the free energy that is responsible for their stability. This is somewhat analogous to chiral noncentrosymmetric magnets, in which the baby-skyrmions are stabilized by the so-called "Dzyaloshinskii-Moriya exchange term." Many additional insights can be obtained from a detailed numerical modeling of the solitonic field configurations through minimization of the Landau–de Gennes free energy supplemented by surface anchoring terms, however, this goes beyond the scope of the present study and will be pursued elsewhere.

To classify the studied field configurations and also to understand them in the context of recent developments in other branches of physics, it is instructive to draw analogies with the two-dimensional skyrmionic structures (baby-skyrmions) in other condensed matter systems. To see this analogy, and for the purpose of simplicity, one can use a vector field to decorate $\hat{n}(\mathbf{r})$. In the LC cell midplane of structures, such as the one shown in Fig. 1, the texture of molecular alignment can be characterized by a topological winding number $N_s$ of the texture commonly referred to as "skyrmion number" and "baryon number" in different branches of science [1–10]. As discussed in the Introduction above, $N_s$ counts the number of times the underlying vector field in the structure wraps a unit sphere. After being decorated by a vector field, the structures shown in Figs. 1–3 would have $N_s$ equal to 1 or −1 (note that the choice of the vector field's directionality is arbitrary due to the nonpolar nature of the LC and the charge can be defined only up to the sign). Along similar lines, when studied in the LC cell midplane, the vector fields decorating structures shown in Figs. 4 and 5 would cover the unit sphere twice, thus having topological skyrmion numbers of $N_s = \pm 2$. Cell midplane textures of field configurations shown in Figs. 8(a)–8(d) can be assigned $N_s$ of $\pm 2$, $\pm 5$, $\pm 9$, and $\pm 3$, respectively (note that the topology of these complex 3D configurations is not fully described by $N_s$ of their 2D field in the cell midplane because the structures are truly 3D in nature). Using an approach illustrated in Fig. 8, in principle, one could generate configurations of arbitrarily large skyrmion number by forming nested loops of CF3 or other cholesteric fingers. It is important to note here, however, that the analogy of studied field configurations with the skyrmionic textures observed in magnetic systems is not complete as our field configurations are truly 3D, a feature that arises due to strong interaction of the molecular alignment field with boundary conditions at the confining surfaces. Having weak surface anchoring boundary conditions (Fig. 3) helps to partly alleviate this effect of confinement. On the other hand, the ability of tuning the strength of surface anchoring interactions along with the nonpolar nature of the used LCs (which allows for existence of topologically stable half-integer defect lines) greatly enrich the diversity of the solitonic field configurations as compared to the ones observed in vector fields and in other condensed matter systems.

The solitonic structures presented in this work can be obtained in both left- and right-handed chiral nematic LCs, as well as when different nematic hosts and chiral agents listed in the materials and techniques section are used, indicating that this structural diversity is not material specific and can be generated in LC systems with different values of elastic constants. We have found that some of the studied laser-generated structures can also be obtained spontaneously when cooling the samples from the isotropic phase (although less reliably and without the control of spatial locations) and after applying low-frequency (10 Hz) electric fields to induce hydrodynamic instability. These localized structures, along with torons and hopfions, are several members of the large family of so-called cholesteric bubbles that were typically studied when occurring spontaneously, but not well understood in past experimental works [20,34–44]. We note that the study of axially symmetric cholesteric configurations reported here is not exhaustive. In fact, several other configurations might have been already identified in prior literature [20,23–26,34–48] that we have not explored so far using our 3D imaging experiments and different types of generation (laser realignment, low-frequency electric field, temperature quench with and without temperature gradients). In addition, in future work, studies of mutually linked multi-toron and multihopfion configurations that also belong to this family will be reported [51].

The study of solitonic configurations in many condensed matter systems is often motivated by the potential of practical applications. For example, the control of individual 2D magnetic baby-skyrmions is a promising breakthrough in the field of spintronics [50,52,53], which may allow for further miniaturization of magnetic memory devices. On the other hand, optically addressable memory devices are of great interest too and our demonstrated laser generation of skyrmionic structures may be of interest for practical optical memory applications. Although individual magnetic skyrmions can be obtained on length scales down to several nanometers, the size of cholesteric localized structures can be potentially tuned down to 100 nm scale. The big advantage is that LC based solitonic structures can be generated and erased by light and electric fields [29–31]. In addition to applications in optically addressed memory devices, electro-optics, singular optics [31], and photonics, our well controlled localized defect configurations can be used in model experiments to entrap nanoparticles of different types, such as metal and semiconductor nanoparticles [30], and to probe their interactions under well defined conditions.

## V. CONCLUSIONS

We have demonstrated controlled optical generation and exploration of a family of localized solitonic configurations, which allow for embedding of axially symmetric twisted structures into confinement-unwound chiral nematic LCs. Three-dimensional optical imaging revealed that most of these embedded solitonic structures can be thought of as looped versions of the so-called cholesteric fingers. Our study demonstrates multiple ways of matching twisted configurations, such as two-dimensional skyrmions of different skyrmion numbers, with the uniform far field and the perpendicular surface boundary conditions. These structures may potentially occur naturally and can be also categorized as subtypes of the generic family of the so-called cholesteric bubbles or spherulitic domains. Because of their topological similarity with various baby-skyrmions and other localized topological field configurations found in other condensed matter systems [49,54], 3D imaging of the director field may translate to better understanding of these other condensed matter systems. On the other hand, the nonpolar LC's ability to host various bulk and surface half-integer disclination loops, in addition to point defects, allows for a realization of configurations that cannot be observed in, for example, chiral magnets. Entrapment of semiconductor and metal nanoparticles in singular defect loops and point defects of these localized structures may enable experiments on these nanoparticles under controlled nanoscale confinement [30,55].

## ACKNOWLEDGMENTS

This work was supported by the Division of Chemical Sciences, Geosciences, and Biosciences, Office of Basic Energy Sciences of the US Department of Energy under Contract No. DE-AC36-08GO28308 with the National Renewable Energy Laboratory (J.v.d.L. and P.J.A.) and the NSF Grants No. DMR-0820579 (R.P.T and I.I.S.) and No.DMR-0847782 (B.S. and I.I.S.). We also thank A. Bogdanov, N. Clark, S. Čopar, J. Evans, J. Fukuda, T. Porenta, C. Twombly, and S. Žumer for discussions.

**References**

[1] T. H. Skyrme, Proc. R. Soc. London Ser. A 260, 127 (1961).

[2] S. Heinze, K. von Bergmann, M. Menzel, J. Brede, A. Kubetzka, R. Wiesendanger, G. Bihlmayer, and S. Blügel, Nat. Phys. 7, 713 (2011).

[3] M. Uchida and A. Tonomura, Nature (London) 464, 737 (2010).

[4] J. Verbeeck, H. Tian, and P. Schattschneider, Nature (London) 467, 301 (2010).

[5] S. Muhlbauer et al., Science 323, 915 (2009).

[6] U. A. Khawaja and H. Stoof, Nature (London) 411, 918 (2001).

[7] J. Fukuda and S. Žumer, Nat. Commun. 2, 246 (2011).

[8] A. N. Bogdanov, U. K. Rößler, and A. A. Shestakov, Phys. Rev. E 67, 016602 (2003).

[9] A. N. Bogdanov and A. A. Shestakov, J. Exp. Theor. Phys. 86, 911 (1998).

[10] U. K. Rößler, A. N. Bogdanov, and C. Pfleiderer, Nature (London) 442, 797 (2006).

[11] R. M. Hornreich and S. Shtrikman, Liq. Cryst. 5, 777 (1989).

[12] G. Heppke, B. Jérôme, H.-S. Kitzerow, and P. Pieranski, Liq. Cryst. 5, 813 (1989).

[13] R. M. Hornreich, M. Kugler, and S. Shtrikman, J. Phys. Colloques 46, C3-47 (1985).

[14] P. Pieranski, P. E. Cladis, and R. Barbet-Massin, J. Phys. Lett. 46, 973 (1985).

[15] R. M. Hornreich, M. Kugler, and S. Shtrikman, Phys. Rev. Lett. 54, 2099 (1985).

[16] W. T. M. Irvine and D. Bouwmeester, Nat. Phys. 4, 716 (2008).

[17] P. Milde, D. Köhler, J. Seidel, L. M. Eng, A. Bauer, A. Chacon, J. Kindervater, S. Mühlbauer, C. Pfleiderer, S. Buhrandt, C. Schütte, and A. Rosch, Science 340, 1076 (2013).

[18] S. Wintz, C. Bunce, A. Neudert, M. Körner, T. Strache, M. Buhl, A. Erbe, S. Gemming, J. Raabe, C. Quitmann, and J. Fassbender, Phys. Rev. Lett. 110, 177201 (2013).

[19] Y. Kawaguchi, M. Nitta, and M. Ueda, Phys. Rev. Lett. 100, 180403 (2008).

[20] P. Oswald, J. Baudry, and S. Pirkl, Phys. Rep. 337, 67 (2000).

[21] L. Gil and J. M. Gilli, Phys. Rev. Lett. 80, 5742 (1998).

[22] I. I. Smalyukh, B. I. Senyuk, P. Palffy-Muhoray, O. D. Lavrentovich, H. Huang, E. C. Gartland, Jr., V. H. Bodnar, T. Kosa, and B. Taheri, Phys. Rev. E 72, 061707 (2005).

[23] A. E. Stieb, J. Phys. 41, 961 (1980).

[24] R. A. Kashanow, J. E. Bigelow, H. S. Cole, and C. R. Stein, Liq. Cryst. Ordered Fluids 2, 483 (1974).

[25] T. Akahane and T. Tako, Mol. Cryst. Liq. Cryst. 38, 251 (1977).

[26] V. G. Bhide, S. Chandra, S. C. Jain, and R. K. Medhekar, J. Appl. Phys. 47, 120 (1976).

[27] F. Simoni and O. Francescangeli, J. Phys.: Condens. Matter 11, R439 (1999).

[28] I. I. Smalyukh, Y. Lansac, N. Clark, and R. Trivedi, Nat. Mater. 9, 139 (2010).

[29] B. G.-g. Chen, P. J. Ackerman, G. P. Alexander, R. D. Kamien, and I. I. Smalyukh, Phys. Rev. Lett. 110, 237801 (2013).

[30] J. S. Evans, P. J. Ackerman, D. J. Broer, J. van de Lagemaat, and I. I. Smalyukh, Phys. Rev. E 87, 032503 (2013).

[31] P. J. Ackerman, Z. Qi, and I. I. Smalyukh, Phys. Rev. E 86, 021703 (2012).

[32] I. I. Smalyukh, S. V. Shiyanovskii, and O. D. Lavrentovich, Chem. Phys. Lett. 336, 88 (2001).

[33] R. P. Trivedi, T. Lee, K. A. Bertness, and I. I. Smalyukh, Opt. Express 18, 27658 (2010).

[34] W. E. L. Haas and J. E. Adams, Appl. Phys. Lett. 25, 535 (1974).

[35] T. Akahane and T. Tako, Jpn. J. Appl. Phys. 15, 1559 (1976).

[36] S. Pirkl, P. Ribiere, and P. Oswald, Liq. Cryst. 13, 413 (1993).

[37] M. Kawachi, O. Kogure, and Y. Kato, Jpn. J. Appl. Phys. 13, 1457 (1974).

[38] S. Pirkl and P. Oswald, J. Phys. II France 6, 355 (1996).

[39] N. Nawa and K. Nakamura, Jpn. J. Appl. Phys. 17, 219 (1978).

[40] W. E. L. Haas and J. E. Adams, Appl. Phys. Lett. 25, 263 (1974).

[41] C. Obayashi, Jpn. J. Appl. Phys. 20, 1753 (1981).

[42] S. Hirata, T. Akahane, and T. Tako, Mol. Cryst. Liq. Cryst. 75, 47 (1981).

[43] V. G. Bhide, S. C. Jain, and S. Chandra, J. Appl. Phys. 48, 3349 (1977).

[44] I. N. Gurova and O. A. Kapustina, Liq. Cryst. 6, 525 (1989).

[45] A. N. Bogdanov, JETP Lett. 71, 85 (2000).

[46] J. M. Gilli and L. Gil, Liq. Cryst. 17, 1 (1994).

[47] A. Bogdanov, JETP Lett. 62, 247 (1995).

[48] U. K. Rößler, A. A. Leonov, and A. N. Bogdanov, Joint Eur. Mag. Symposia 303, 012105 (2010).

**Figures**

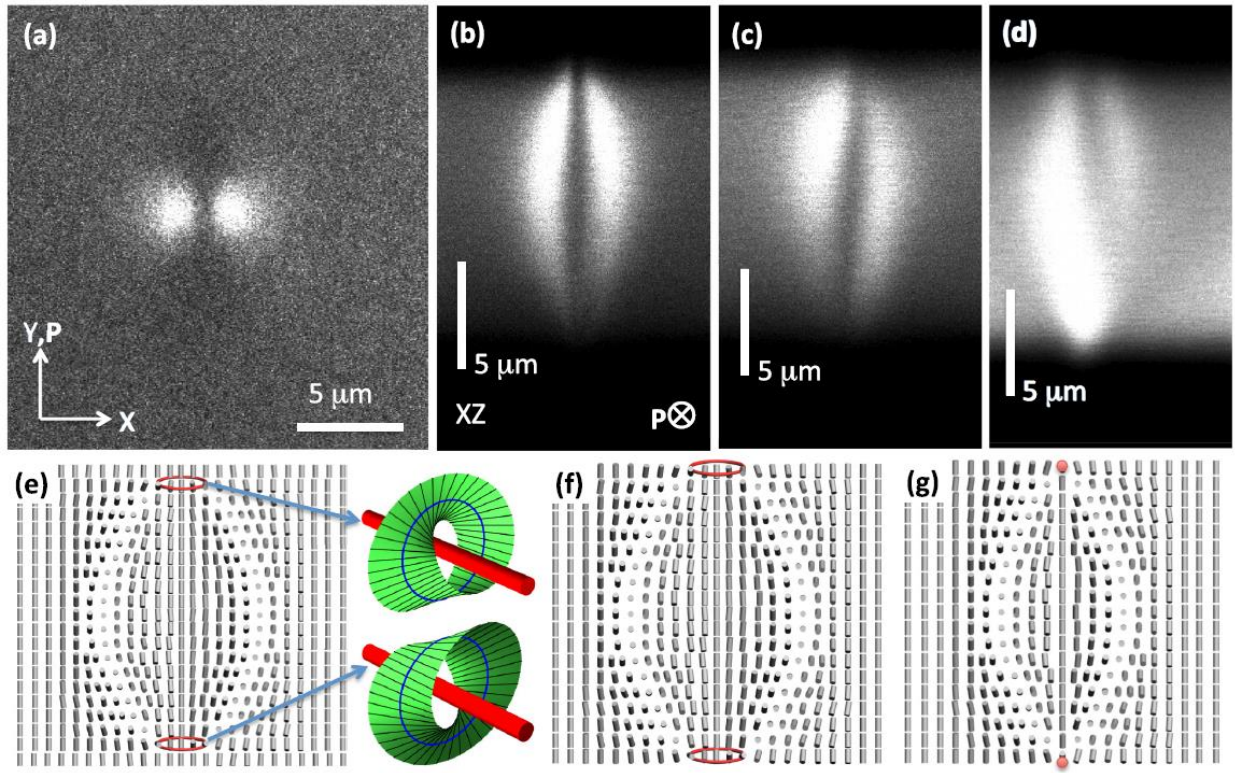

FIG. 1. Baby-skyrmion localized cholesteric structures. (a) Depth-resolved FCPM image obtained in the midplane of the LC cell. (b)–(d) Vertical cross sections of several slightly different baby-skyrmion structures obtained using FCPM linear polarizations "P" normal to the images, as depicted in (b). (e)–(g) Schematics of molecular alignment configurations reconstructed for (e) strong and (f) and (g) relatively weak surface anchoring boundary conditions leading to the formation of (e) and (f) twist disclination loops (e) in the LC bulk and (f) at surfaces as well as (g) surface point defects called "boojums." The twist-disclination loops are shown in red (dark gray) and the director field structures around them are depicted on Möbius strips. The structure was realized in a cell of $d \approx 10$ μm filled with a cholesteric mixture formed from the nematic host ZLI-2806 and chiral additive CB15.

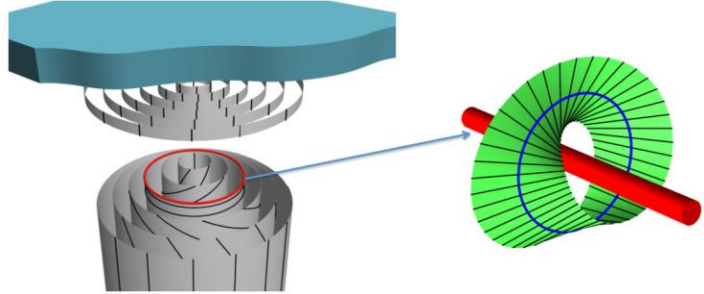

FIG. 2. A twist disclination loop [depicted as a red (gray) line] allows for matching the field configuration of a double twist cylinder with the strong vertical surface boundary conditions. The director field on concentric circles around the defect line circumscribes the surface of a Möbius strip, as shown in the inset.

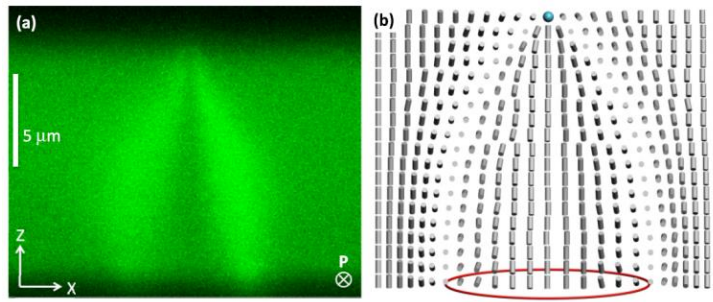

FIG. 3. A localized axially symmetric field configuration arising in a cell with weak perpendicular surface anchoring. (a) FCPM vertical cross section obtained for a linear polarization "P" of probing light being orthogonal to the image. (b) The corresponding director structure depicting a surface boojum (blue sphere) on one of the substrates and a surface disclination [red (gray) loop] on the other confining substrate. The structure was realized in a cell of $d \approx 10$ μm filled with a cholesteric mixture formed from the nematic host ZLI-2806 and chiral additive CB15.

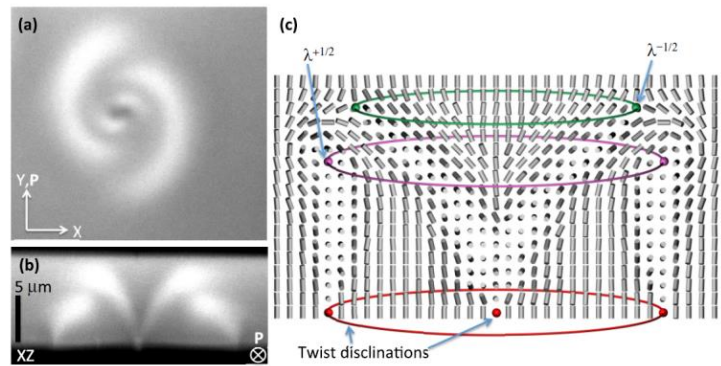

FIG. 4. An axially symmetric structure formed by loops of nonsingular $\lambda$ disclinations and twist disclinations interspaced by twist. (a) and (b) The structure is imaged using FCPM with linearly polarized excitation light in the (a) in-plane and (b) vertical cross sections of the sample. (c) Schematic representation of the axially symmetric structure. The structure contains two nonsingular $\lambda$-disclination loops in the bulk and two singular twist-disclination loops at one of the confining surfaces; one of the disclination loops appears to be shrunk to a point defect, which might be actually a small disclination ring that cannot be resolved in the image due to the limited resolution of the FCPM technique. The structure was realized in a cholesteric cell of $d \approx 10$ μm filled with a mixture of nematic host ZLI-2806 and chiral additive CB15.

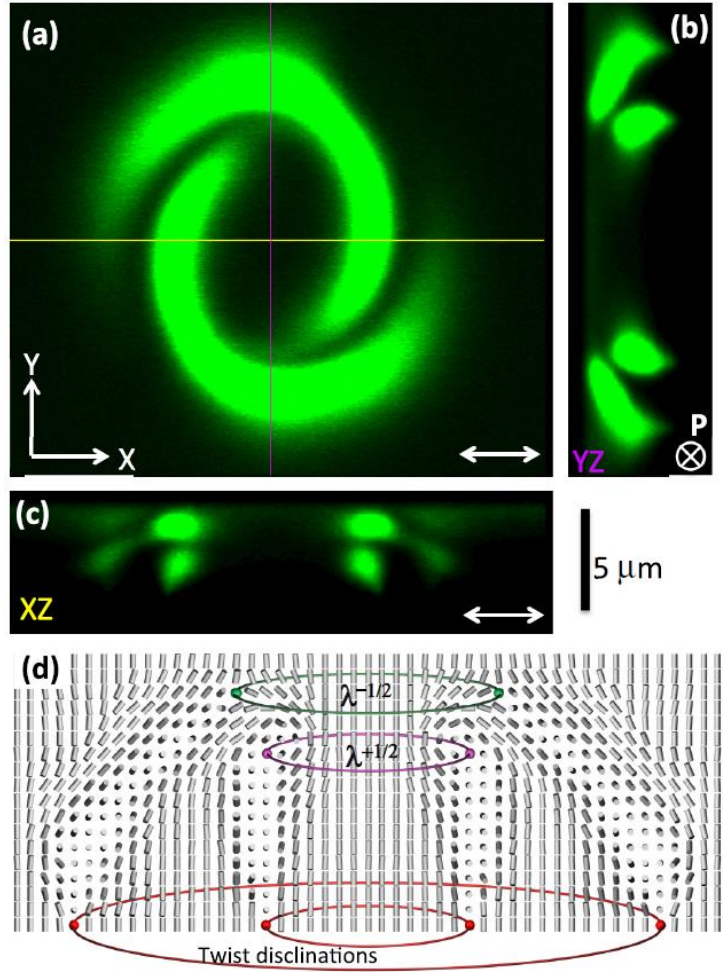

FIG. 5. A localized twisted configuration with $2\pi$ twist in all radial directions matched to the perpendicular surface boundary conditions by two nonsingular $\lambda$ disclinations and two twist disclinations. (a) In-plane 3PEF-PM image obtained for cell midplane and linear polarization direction of the excitation light marked on the image by a double arrow. (b) A vertical 3PEF-PM cross section along the magenta (dark gray) line marked in (a) and for the 3PEF-PM excitation light's polarization direction normal to the image, as marked by a white crossed circle. (c) A vertical 3PEF-PM cross section obtained along the yellow (light gray) line shown in (a) and for a polarization direction marked by the white double arrow. (d) A schematic showing the axially symmetric director structure with the $2\pi$ twist in all radial directions and four disclination loops matching this twist to the perpendicular surface boundary conditions at the confining substrates. The images were obtained using cells with substrates treated with DMOAP and separation of 5 μm filled with cholesteric LC made of a nematic E7 and chiral dopant CB15.

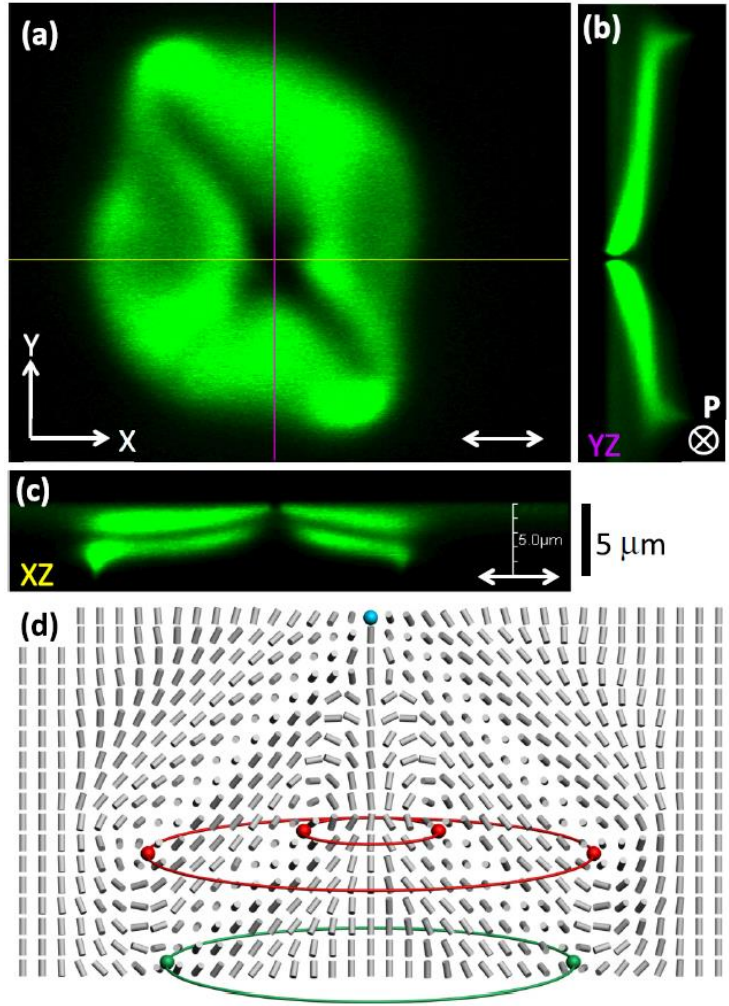

FIG. 6. A localized axially symmetric twisted configuration with the helical axis tilted with respect to cell substrates and this twist matched to the perpendicular surface boundary conditions by a boojum (blue circle) and disclination loops (red $\lambda^{+1/2}$ and green $\lambda^{-1/2}$ circles). (a) In-plane 3PEF-PM micrograph obtained for the cell midplane and linear polarization direction of the excitation light marked on the image by a white double arrow. (b) A vertical 3PEF-PM cross section along the magenta (dark gray) line marked in (a) and for the 3PEF-PM excitation light's polarization direction normal to the image, as marked by a white crossed circle. (c) A vertical 3PEF-PM cross section obtained along the green (light gray) line shown in (a) and for a polarization direction marked by the white double arrow. (d) A schematic showing the axially symmetric director structure with the $2\pi$ twist in all radial directions and disclination loops and boojum matching this twist to the perpendicular surface boundary conditions at the confining substrates. The images were obtained using cells with substrates treated with DMOAP and separation of 5 μm filled with cholesteric LC made of E7 and chiral dopant CB15.

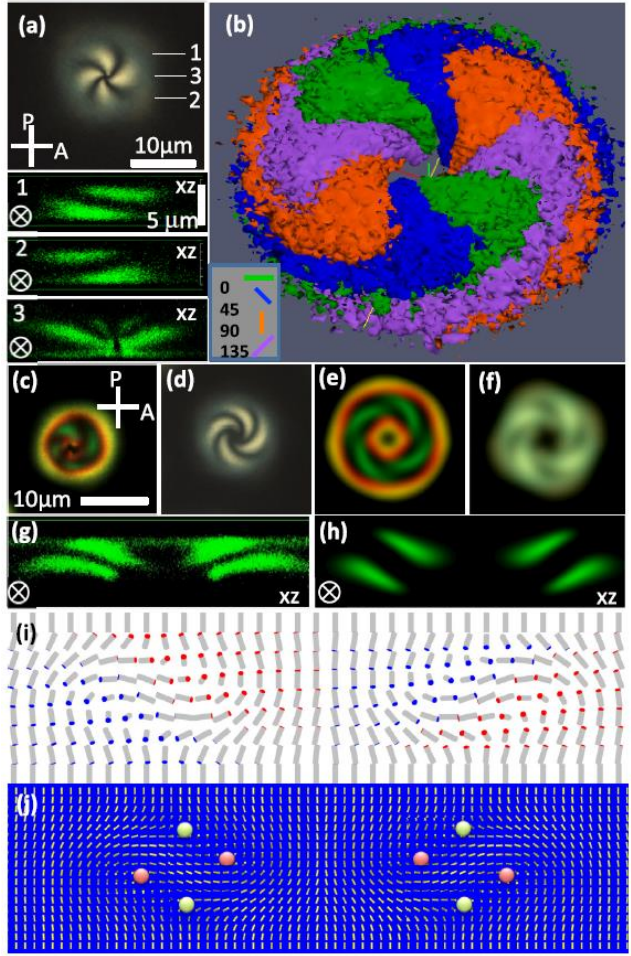

FIG. 7. Cholesteric localized structures of finger loops (a) and (b) with one nonsingular $\lambda$ disclination shrunk into a point defect (c)–(i) or not. (a) A POM micrograph obtained between crossed polarizers with three 3PEF-PM vertical *x-z* cross sections marked on the image and shown beneath it. The linear polarization direction of the used excitation light is perpendicular to the planes of the 3PEF-PM vertical cross sections (1–3), as marked by crossed circles. (b) A ParaView presentation of the 3D director structure shown in (a). The perspective view is from the side of the point defect formed by shrinking a nonsingular $\lambda$ disclination loop. The image was constructed by superimposing high-intensity regions of 3D 3PEF-PM textures obtained for four different linear polarization directions depicted in the bottom-left inset. (c)–(f) POM optical micrographs (c) and (d) obtained experimental and (e) and (f) by use of computer simulations for (c) and (e) high-birefringence LC and (d) and (f) low-birefringence LC. (g) and (h) 3PEF-PM vertical cross sections obtained (g) experimentally and (h) by use of computer simulations for a minimum-energy director structure shown in (i). (i) Axially symmetric director structure of a looped CF1 finger obtained by means of computer simulations via minimization of the elastic free energy given by Eq. (1) and assuming infinitely strong surface anchoring boundary conditions. Red and blue (dark and light gray) cylinder caps illustrate orientation of the vector-field-decorated **n(r)** (pointing into or out of the page). (j) A schematic equivalent to that shown in (i) but with more detailed presentation of the structure using a large number of smaller cylinders depicting changes in orientation of $\hat{n}(\mathbf{r})$ and circles representing locations of disclination loops of $\lambda^{+1/2}$ [red (dark gray)] and $\lambda^{-1/2}$ [green (light gray)]. The images were obtained using wedge cells formed from polyimide-treated glass plates with thickness varying in the range of 7–10 μm filled with partially polymerizable cholesteric LC composite described in the materials section.

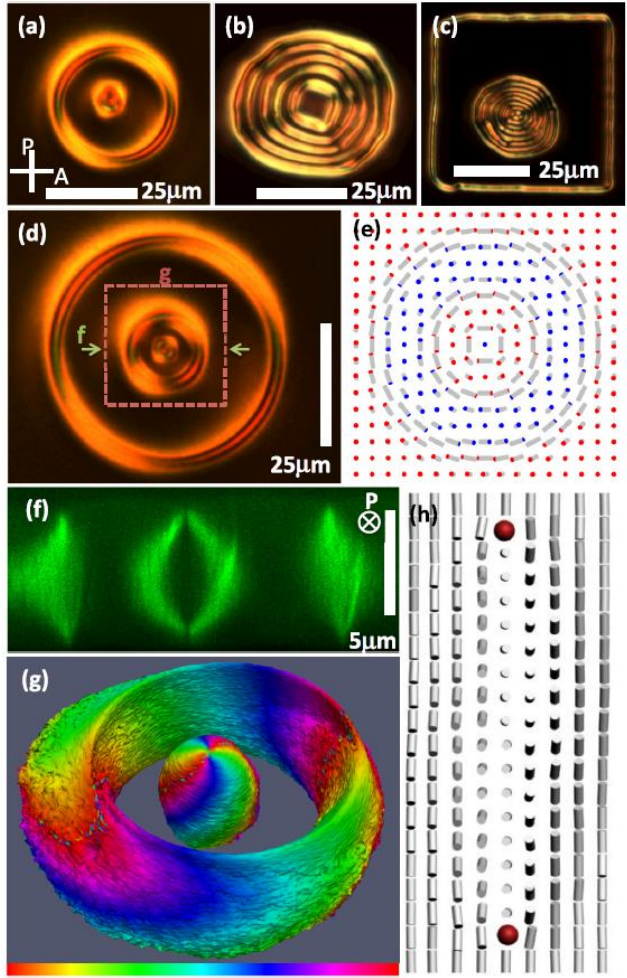

FIG. 8. Localized director structures with large topological skyrmion numbers. (a)–(d) POM micrographs of laser-generated configurations having $2\pi$, $5\pi$, $9\pi$, and $3\pi$ twist in the midplane of the cell in all radial directions. (e) Schematic of the axially symmetric midplane cross section of (d) showing the $3\pi$ twist from center to periphery in all radial directions. Red (dark gray) and blue (light gray) cylinder caps illustrate orientation of the vector-field-decorated n(r) (pointing out of or into the page). (f) 3PEF-PM cross section of the structure along the f-f line marked in (d). (g) ParaView presentation of the 3D director structure within the dashed square region of (d) with a toron in the center containing point defects being surrounded by a concentric CF3 finger loop. The color (grayscale levels) encodes the azimuthal molecular/director orientation (molecular orientation changing by 180 deg clockwise around the axis of symmetry corresponds to transitioning from red to green and then to blue and, finally, back to red). (h) A schematic of the director field configuration of the CF3 finger used to form nested loops shown in (a)–(d). The singular twist disclinations within the CF3 finger's cross section are marked by red (dark gray) filled circles. The images in (a), (d), (f), and (g) were obtained using wedge cells formed from polyimide-treated substrates with thickness varying in the range of 7–10 μm filled with partially polymerizable cholesteric LC composite described in the materials section. The images in (b) and (c) were obtained using similar wedge cells formed from polyimide-treated glass substrates filled with a cholesteric LC of pitch equal to 8 μm formed from a mixture of E7 and CB15.